\input{aipcheck}

\documentclass[
    ,final            
  ]
  {aipproc}

\layoutstyle{6x9}

\begin{document}

\title{Setting
the scale for DIS at large Bjorken x}


\author{Simonetta Liuti}{
  address={Department of Physics, University of Virginia, Charlottesville, VA 22904, USA}
}

\begin{abstract}
We discuss the extension of a systematic perturbative QCD based analysis
to the $x \rightarrow 1$ region.  
After subtracting a number of effects that transcend NLO pQCD evolution, 
such as target mass corrections and large $x$ resummation effects, 
the remaining power corrections can be interpreted as dynamical higher twists. 
The quantitative outcome of the analysis is dominated by the interplay between the value of $\alpha_S$ in the 
infrared region and the higher twists. We uncover a dual role played by $\alpha_S$ at large Bjorken $x$ 
that can be used to experimentally extract its 
value in the non-perturbative regime. 
\end{abstract}

\maketitle


\section{Introduction}
\begin{quote}
{\it ``QCD nowadays has a split personality. It embodies 'hard' and 'soft' physics, 
both being hard subjects and the softer the harder.'' } \cite{Dok}
 \end{quote}
 
QCD's main goal is to describe the structure 
of hadrons in terms of its fundamental degrees of freedom, the quarks and gluons (partons). Hadrons are observable  
in both the initial and final stages of hard processes, 
while the existence and properties of partons are inferred only indirectly. 
The underlying idea is that because of the smallness of the coupling constant at large enough momentum transfer, $Q^2$, 
or equivalently at short distances ${\cal O}(1/\sqrt{Q^2})$, 
a hard probe sees hadrons as composed of "free" quarks and gluons carrying fractions 
$x$ of the hadron's Light Cone (LC) momentum, with given probability distributions. In  Deep Inelastic Scattering (DIS)
the latter are identified with the Parton Distribution Functions
(PDFs)  \cite{CTEQ_Handbook}.  
As increasingly shorter distances are probed the PDFs shape in $x$ changes due to radiative processes,  according to a pattern which is 
calculable with high accuracy within Perturbative QCD (PQCD) \cite{Moch}. 
Factorization theorems regulate this fundamental property of the theory thereby allowing for
the short distance  behavior
to be evaluated using as input the values
taken by the PDFs at a given initial $Q^2$. 
The strong coupling regime of QCD remains incalculable; it is accessible only with 
non-perturbative (approximate) methods.
 
The separation and yet coexistence of long distance and short distance structure in QCD
has by now become naturally accepted as part of a ``common wisdom framework'' underlying the interpretation of most 
experiments, from Deep Inelastic Scattering (DIS) 
to $e^+e^- \rightarrow$ hadrons, to hadron-hadron scattering. 
The concept of {\em duality} is implicitly used to various degrees, 
 meaning, in the most extreme case, that hadronic 
observables are replaced by calculable partonic ones with
little more going into the hadronic formation phase 
of the process (from partons to hadrons or vice versa). 
In a phenomenological context, duality purports 
to study how a number of properties
defined from the beginning of the hard scattering process, 
are predetermined and persist in the non-perturbative stage.
A particularly striking realization of duality, known as Bloom Gilman duality \cite{BG}, is observed in DIS, where for 
large values of Bjorken $x >0.5$ ($x=Q^2/2M\nu$, $M$ being the proton mass and $\nu$ 
the energy transfer in the lab system), and for $Q^2 \approx$ 5 GeV$^2$, one has 
an invariant mass of $W^2 \leq 5$ GeV$^2$ ($W^2 = Q^2(1/x-1)+M^2$), 
lying  mostly in the resonance region. While it is impossible to reconstruct the detailed structure of the
proton's resonances, these remarkably follow the PQCD predictions when averaged over  $x$ (see {\it e.g.} \cite{Mel} for a review).

Duality in inclusive scattering is clearly a phenomenological manifestation of the non-perturbative to perturbative transition in QCD, whose
origins are still largely unknown.
It is now becoming mandatory to have a fuller understanding of its working, motivated on one side by the existence of highly  accurate data at large $x$, and spurred, on the other, by the advent of the LHC 
where previously unexplored regimes in $x$ and $Q^2$ will be within reach.        

Our studies of the physical origin of duality, and of its impact on our understanding of the nucleon's structure started a few years ago when we 
set up a program to quantitatively extract the scale dependence of the large $x$ Jlab Hall C data \cite{Ioana}. In the analysis performed in Refs.\cite{Kep1,Kep2} 
we addressed several effects that have a large impact at large $x$, namely Target Mass Corrections (TMCs), large $x$ resummation effects, and 
higher twists.
This work  was then completed, and extended to polarized data in Ref.\cite{BFL}. 
An important point emerged from Refs.\cite{Kep1,Kep2,BFL} that a deeper understanding was needed 
of those aspects unique to the large $x$ perturbative QCD analysis. 
Our point of view is that only after a  complete perturbative QCD analysis  is performed can one {\em define} duality by  quantitatively establishing
whether, and to what extent, this phenomenon is responsible for the apparent cancellation of multiparton correlations.  
This point of view is somewhat complementary to the approaches of  Refs.\cite{CloIsg,CloMel,Jes}, which
are quark model based and focused on symmetry aspects. 

In this contribution we argue that once the range of validity of parton-hadron duality is defined quantitatively, 
a precise PQCD analysis at  large $x$ would open up the possibility of extracting 
the strong coupling constant, $\alpha_S$, at low scale. 
Such an analysis would complement the 
recent extractions using data on the GDH sum rule \cite{ChenDeur}. It would, furthermore, add insight on a recent interpretation put forth in Ref.\cite{Deur1} of
the effective coupling constant in the strongly interacting/non perturbative regime from light front holographic mapping of classical gravity in Anti de-Sittter (AdS) space.

In order to explain our approach, we first present an overview of the large $x$ data, and discuss a few aspects of the evolution mechanism for 
DIS at large $x$ where two scales related to the invariant mass
and to the four-momentum transfer, are simultaneously present. We reiterate that accurate analyses in this region  
such as the ones first conducted {\it e.g.} in \cite{Kep1,Kep2,BFL} are crucial for establishing the interplay of the various components ($\alpha_S$, multiparton correlations, etc..) of the perturbative to non-perturbative transition regime in QCD. We then illustrate the connection between large $x$ data 
and $\alpha_S$ in the infrared region, and draw our conclusions.

\section{Analysis of large $x$ data}
High precision inclusive unpolarized electron-nucleon scattering 
data on both hydrogen and deuterium targets from Jefferson Lab are available 
to date in the large $x$, multi-GeV regime (see \cite{Christy} and references therein).
Because of the precision of the data one should now be able to distinguish among  different sources of 
scaling violations affecting the structure functions in addition to standard NLO evolution, 
\begin{itemize}
\item Target Mass Corrections
(TMC), 
\item Large $x$ Resummation Effects (LxR)
\item Nuclear Effects 
\item Dynamical Higher Twists (HTs), 
\item Impact of Next-to-Next-to-Leading-Order (NNLO) perturbative
evolution. 
\end{itemize}
All of the effects above can be extracted with an associated theoretical error. It is in fact well known that their evaluation is model dependent.
Recent studies, however, have been directed at determining more precisely both the origin and size of the associated theoretical error. 
Recent analyses have been taking into account, so far, some but not all of the effects listed above \cite{Acc_10}. 

\subsection{Unpolarized structure function.}
The inclusive DIS cross section of unpolarized electrons off
an unpolarized proton is written in terms of the two structure
functions $F_2$ and $F_1$, 
\begin{eqnarray}
\label{xsect}
\frac{d^2\sigma}{dx dy} =
\frac{4\pi\alpha^2}{Q^2 xy}
\left[
    \left(1-y-\frac{(Mxy)^2}{Q^2}\right)F_2 +
    y^2 x F_1    \right], 
\end{eqnarray}
with $y=\nu/\epsilon_1$, $\epsilon_1$ being the initial electron energy.
The structure functions are related by the equation  
\begin{equation}
\label{R}
F_1 = F_2(1+\gamma^2)/(2x(1+R)),
\end{equation}
where $\gamma^2=4M^2x^2/Q^2$, and $R$ is ratio of the longitudinal to transverse
virtual photo-absorption cross sections. 
%
In QCD, $F_2$ is expanded in series of inverse powers of $Q^2$, 
obtained by ordering the matrix 
elements in the DIS process by increasing twist $\tau$, which is equal
to their dimension minus spin
\begin{eqnarray}
\label{t-exp} 
F_{2}(x,Q^2)  =  F_{2}^{LT}(x,Q^2) +
\frac{H(x)}{Q^2} + {\cal O}\left(\frac{1}{Q^4} \right) \simeq F_{2}^{LT}(x,Q^2) \left(
1+ \frac{C(x)}{Q^2} \right) + {\cal O}\left(\frac{1}{Q^4} \right) 
\end{eqnarray}
The first term is the leading twist (LT), with $\tau=2$.
The terms of order $1/Q^{\tau-2}$, $\tau \geq 4$, in Eq.(\ref{t-exp}) 
are the higher order terms, generally referred to as 
higher twists \cite{CTEQ_Handbook}. 

\newpage 
\centerline{\it Target Mass Corrections} 
\vspace{0.5cm}

TMCs are included in $F_2^{LT}$.
For $Q^2$ $\geq$ 1 GeV$^2$, TMCs can be taken into
account through the following expansion \cite{DGP} 
\begin{eqnarray}
\label{TMC}
F_{2}^{LT(TMC)}(x,Q^2)  = 
    \frac{x^2}{\xi^2\gamma^3}F_2^{\mathrm{\infty}}(\xi,Q^2) + 
    6\frac{x^3M^2}{Q^2\gamma^4}\int_\xi^1\frac{d \xi'}{{\xi'}^2} 
F_2^{\mathrm{\infty}}(\xi',Q^2),
\end{eqnarray}
where $F_2^{\infty}$
is the structure function in the absence of TMCs. A more recent analysis \cite{AccQiu}
re-examined TMCs within the collinear factorization approach of \cite{ColRogSta} in order to address the 
longstanding question of  
the unphysical behavior in the threshold region of Eq.(\ref{TMC}). This originates 
from the fact that as $x \rightarrow 1$, one  
obtains a $Q^2$-dependent threshold, namely $F_2(\xi,Q^2) =0$ for  $\xi > \xi_{max} = 2/1+\sqrt{1+ 4 M^2/Q^2}$, therefore 
rendering $F_2$ undefinable as $Q^2$ varies (see discussion in \cite{Schien}).
In the formalism of Ref.\cite{AccQiu}, TMCs, applied to the helicity dependent structure functions read
\begin{equation}
F_{T}^{LT(TMC)}(x,Q^2) = \int^{\frac{1-x}{x}Q^2}_{m_\pi^2}  d m_J^2 \rho(m_J^2) F_T^\infty\left[ \xi \left(1+\frac{m_J^2}{Q^2}\right),Q^2 \right], 
\end{equation} 
where $F_T \equiv F_1$. The final quark is assumed to hadronize into a jet of mass $m_J$ with a  a process dependent distribution/smearing function $\rho(m_J^2)$.
In our extraction we take the perspective that the evaluation of TMCs is always associated with  
the evaluation of HTs --
TMCs should in principle be applied also to HTs --  in an inseparable way. Therefore  
we consistently keep terms of ${\cal O}(1/Q^4)$ \cite{BFL,AKL}, whether in the formalism/prescription of Ref.\cite{DGP} or of Ref.\cite{AccQiu}.
$H(x,Q^2)$, then, represents the ``genuine'' HT correction that involves
interactions between the struck parton and the spectators or, formally,
multi-parton correlation functions.   
%

\vspace{0.5cm}
\centerline{\it Threshold Resummation} 
\vspace{0.5cm}

In order to understand the nature of the remaining $Q^2$ dependence that cannot
be described by NLO pQCD evolution, we also include the effect of  
threshold resummation, or Large $x$ Resummation (LxR).
LxR effects arise formally from terms containing powers of 
$\ln (1-z)$, $z$ being the longitudinal 
variable in the evolution equations, that are present in 
the Wilson coefficient functions $C(z)$. 
Below we write schematically how the latter relate the parton distributions to {\it e.g.} 
the structure function $F_2$, 
\begin{equation}
F_2^{LT}(x,Q^2)  = \frac{\alpha_s}{2\pi} \sum_q \int_x^1 dz \, C(z) \, q(x/z,Q^2), 
\label{lxr}
\end{equation}   
where we have considered only the non-singlet (NS) contribution to $F_2$ since 
only valence quarks distributions are relevant in our kinematics. 
The logarithmic terms in $C(z)$ become very large at large $x$, and they need to be 
resummed to all orders in $\alpha_S$. 
Resummation was first introduced by  
linking this issue to the definition of the correct kinematical variable that determines the 
phase space for the radiation of gluons
at large $x$. This was found to be $\widetilde{W}^2 = Q^2(1-z)/z$, 
instead of $Q^2$ \cite{BroLep,Ama}. 
As a result, the argument of the strong coupling constant becomes $z$-dependent: 
$\alpha_S(Q^2) \rightarrow \alpha_S(Q^2 (1-z)/z)$ \cite{Rob,Rob1}. 
In this procedure, however, an ambiguity is introduced, related to the need of continuing 
the value of $\alpha_S$  
for low values of its argument, {\it i.e.} for $z \rightarrow 1$ \cite{PenRos}. 
Although on one side, the size of this ambiguity could be of the same order of the HT corrections 
and, therefore,  a source of theoretical error, on the other by performing an accurate analysis such as 
the one proposed here, one can extract $\alpha_S$ for values of the scale in the infrared region. 
We address this point in more detail in the next Section.

\vspace{0.5cm}
\centerline{\it Nuclear Effects} 
\vspace{0.5cm}
Theoretical uncertainties in the deuteron are taken routinely into account, and are expected
to be in sufficient control (see \cite{KulPet} and references therein).
Uncertainties arise mainly from  

\noindent 
{\it i)} Different models
of the so called nuclear EMC effect; 

\noindent 
{\it ii)} Different
deuteron wave functions derived from currently available NN potentials,
giving rise to different amounts of high momentum components;

\noindent
{\it iii)} The interplay between nucleon off-shellness and TMC in nuclei.

\vspace{0.3cm}
Finally, we did not consider NNLO calculations, these are not expected
to alter substantially our extraction since,
differently from what seen originally in the case of $F_3$, these have been proven 
to give a relatively small contribution to $F_2$.

\vspace{0.3cm}  
Once all of the above effects have been subtracted from the data, and assuming the validity 
of the twist expansion, 
Eq.(\ref{t-exp}) in this region, one can 
interpret more reliably any remaining discrepancy in terms of HTs. 
\begin{figure}
  \includegraphics[height=.35\textheight]{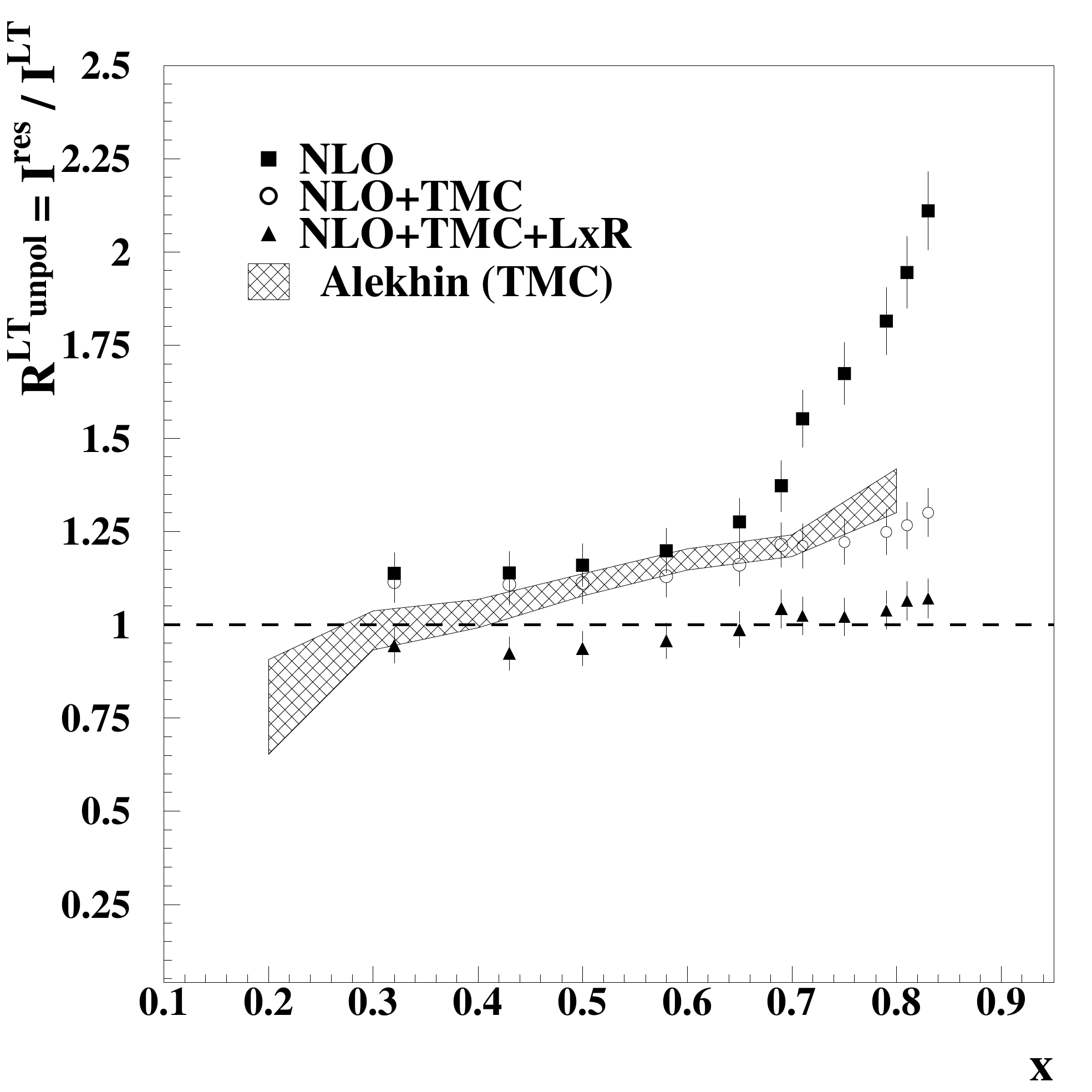}
  \caption{HT coefficients extracted in the resonance region according 
         to the procedure described in the text.  
         HT extracted with only the NLO calculation (squares); the 
         effect subtracting TMC (open circles); the effect of subtracting both 
         TMC and LxR (triangles). Shown for comparison are the values obtained from 
         the coefficient $H$ obtained in Ref.\protect\cite{Ale1} using DIS data and 
         including the effect of TMC.  Adapted from Ref.\cite{BFL}}
\end{figure}
Since we extend our $x$-dependent analysis to the resonance region
we consider the following integrated quantities
\begin{equation}
\label{Iexp}
I^{\mathrm{res}}(\langle x \rangle, Q^2) = \int^{x_{\mathrm{max}}}_{x_{\mathrm{min}}} 
F_2^{\mathrm{res}}(x,Q^2) \; dx
\end{equation}
where $F_2^{\mathrm{res}}$ is evaluated using the experimental data  
in the resonance region. 
For each $Q^2$ value:
$x_{\mathrm{min}}=Q^2/(Q^2+W_{\mathrm{max}}^2-M^2)$, and 
$x_{\mathrm{max}}=Q^2/(Q^2+W_{\mathrm{min}}^2-M^2)$, where 
$W_{\mathrm{min}}$  and $W_{\mathrm{max}}$ delimit the resonance region,
and $\langle x \rangle$ is the average 
value of $x$ for each kinematics. 
This procedure replaces a strict
point by point in $x$, analysis.

Typical results from the analysis outlined above are plotted in Figure 1 where we show the HT coefficient  defined from Eq.(\ref{t-exp}) as
\[ R^{LT} \equiv C(x) = Q^2 \left[F_{2}(x,Q^2)/F_{2}^{LT}(x,Q^2) -1 \right]
  \]
The error in the figure is from the experimental data. No theoretical uncertainty was included.  However, our results clearly show that the combined effects of 
TMCs and LxR substantially reduce $C(x)$. We take this as illustrative of the accomplishments one can expect from the analysis we suggest in this contribution.
Essential features that emerge are the interplay between the values of $\alpha_S(M_Z^2)$ and the HTs, the relevance of TMCs, and, most importantly,  the need to
define $\alpha_S$ in the infrared region.  All of these features can affect the central values of the HTs reported in Fig.1.

\section{$\alpha_S$ at $x \rightarrow 1$}
\begin{figure}
\label{fig2}
  \includegraphics[height=.3\textheight]{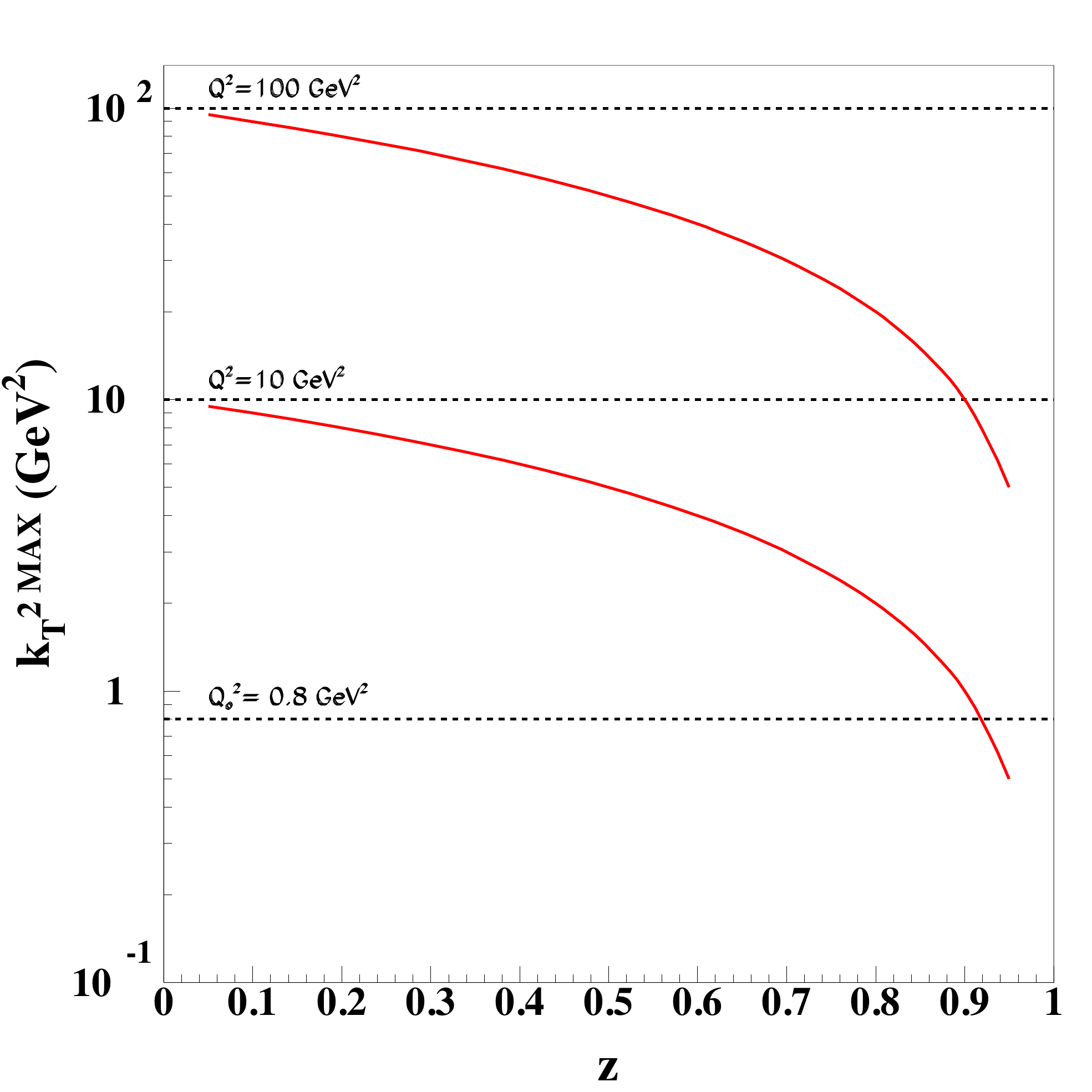}
  \caption{Phase space in the evolution of the NS component. The dotted lines are for $k_{T, \, MAX}^2 = Q^2$, for $Q^2=0.8, 10, 100$ GeV$^2$. 
  The full lines represent the upper limit $k_{T, \, MAX}^2 = Q^2(1-z)/z$ for $Q^2=10, 100$ GeV$^2$. In this case, one can see a clear reduction 
  of the allowed $k_T$ at large $z$.}
  \end{figure}

\begin{figure}
\label{fig3}
  \includegraphics[height=.35\textheight]{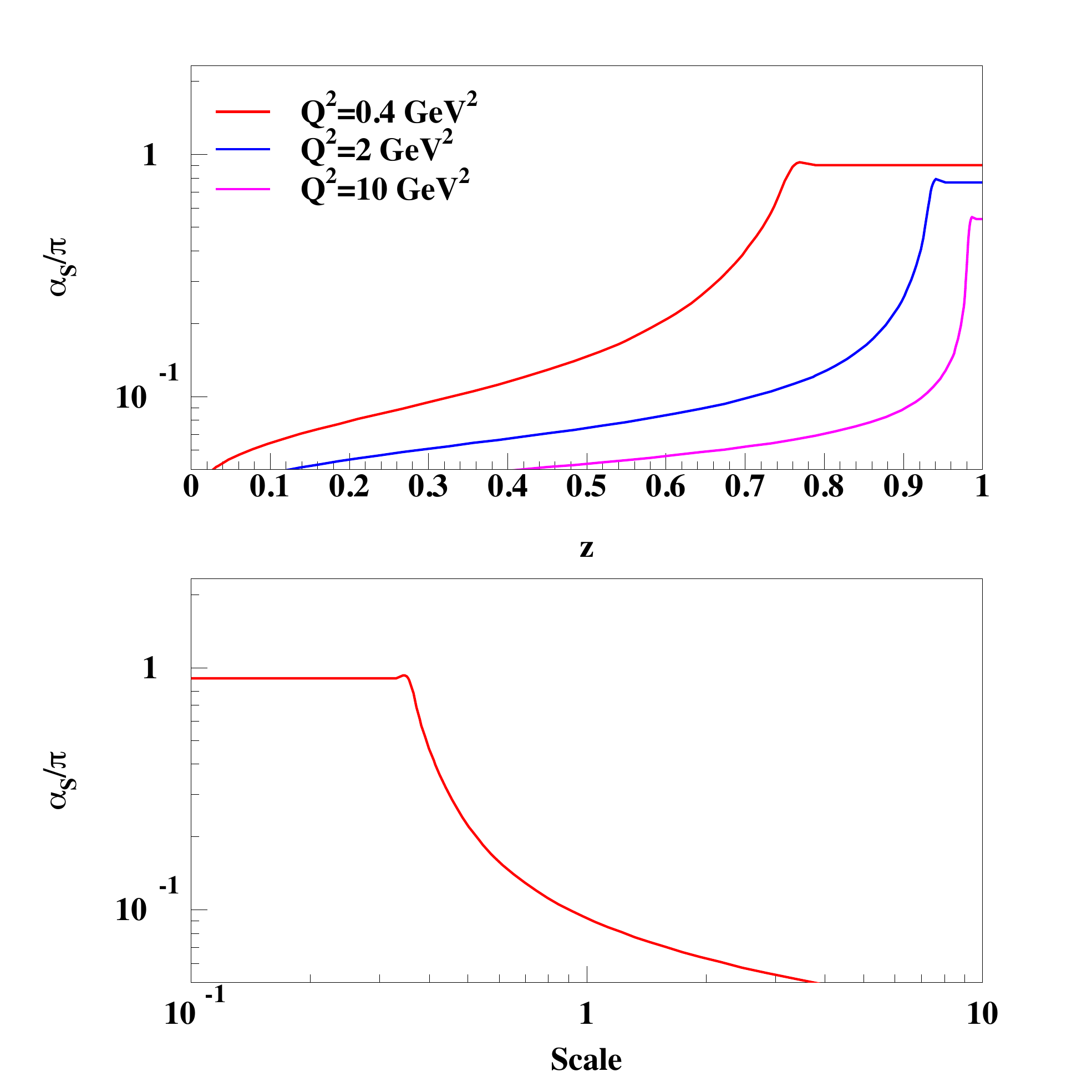}
  \includegraphics[height=.40\textheight]{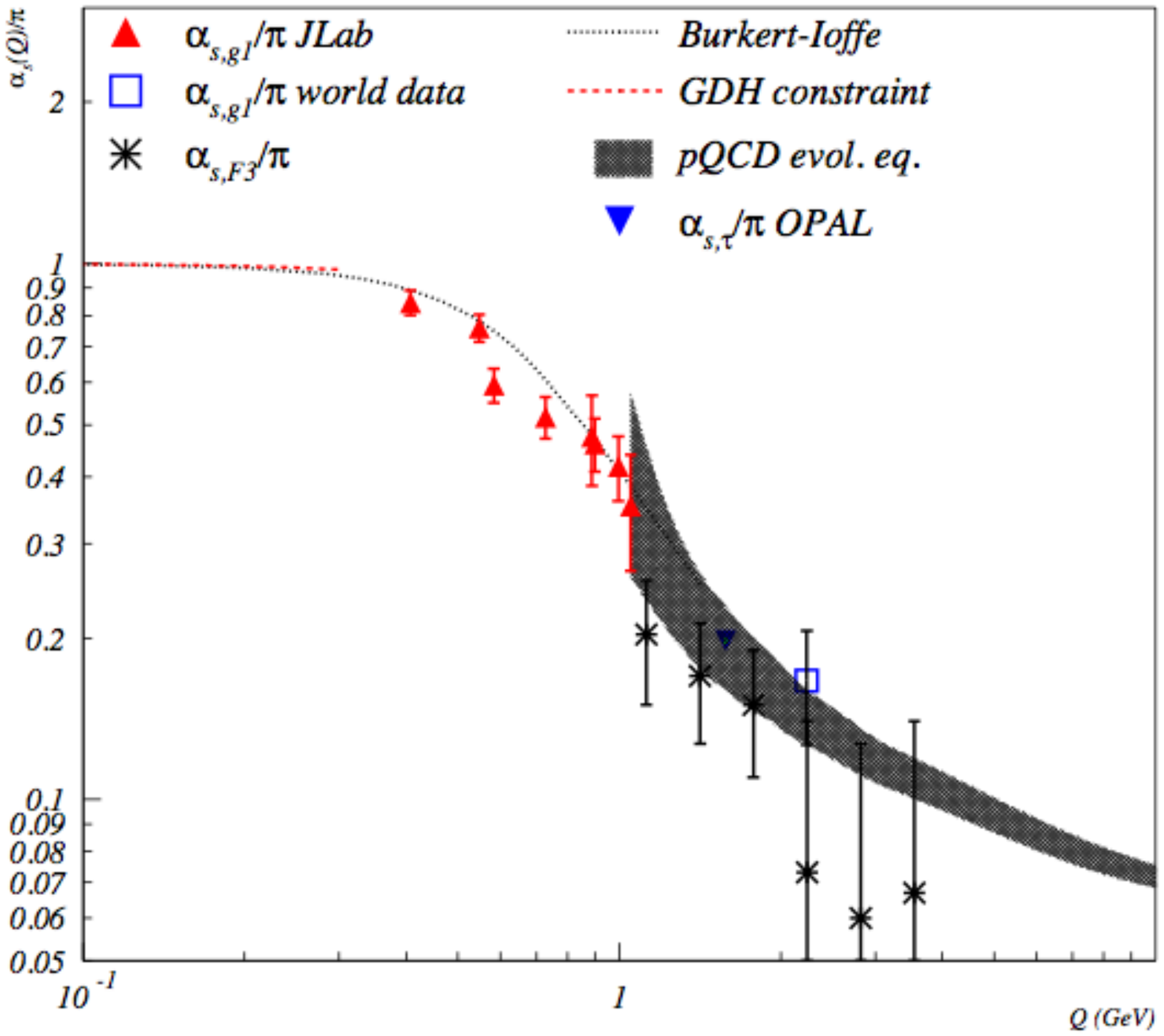}
  \caption{Left panel: $\alpha_S/\pi$ extracted from the analysis of the large $x$ data discussed in the text, and plotted
  vs. $z$ (Eq.(\ref{lxr}) (upper panel), and the scale $\widetilde{W} = \sqrt{Q^2(1-z)/z}$ (lower panel). For comparison 
  we show the extraction from Ref.\cite{ChenDeur} using Jefferson Lab data at $Q^2 = 0.7-1.1$ GeV$^2$.}
  \end{figure}

We now discuss in more detail the working of threshold resummation, and its possible impact 
on the analysis of $F_2$ at large $x$  \cite{Rob}. Starting from NLO, the coefficient, $C(z)$ in Eq.(\ref{lxr}) 
is dominated at large $x$ by terms proportional to $[\alpha_S(Q^2) \ln(1-z)]^n$ which need to be resummed
in the perturbative series. The physical origin of these terms is in the phase space for the contribution
of gluons emission to evolution, which become soft as $x\rightarrow 1$. A mismatch in the cancellation
with the virtual gluons contributions ensues.  
If, however, one carefully evaluates the kinematics for gluon emission at large $x$ 
within a quark-parton model view, one obtains \cite{BroLep}
\begin{eqnarray}
q(x,Q^2) & = & q(x,Q_o^2) + \int_{Q_o^2}^{\widetilde{W}^2} \frac{dk_\perp^2}{k_\perp^2}  
\frac{\alpha_S(k_\perp^2)}{2\pi}
\int_x^1 \frac{dz}{z} P_{qq}\left( \frac{x}{z},
\alpha_S \right) q(z,k_\perp^2),
\end{eqnarray}
where $Q^2_o$ is an arbitrary initial scale, and $\widetilde{W}^2= Q^2(1-z)/z$ is the maximum $k_\perp^2$ in the virtual photon-quark
center of mass system, appearing in the ladder graphs that define the leading log result. The resulting phase space is shown for
different $Q^2$ values in Fig.\ref{fig2}. 
The reduction of the allowed $k_\perp^2$ results  in a simultaneous shift in the argument of $\alpha_S \rightarrow \alpha_S(Q^2/(1-z)/z)$,
and a cancellation of the $\alpha_S(Q^2) \ln(1-z)$ divergence in the NLO coefficient function.  
As a consequence of rescaling the argument of $\alpha_S$ one has to consider its continuation into the infrared region \cite{Rob1,BFL}. 
The left panels of Fig.\ref{fig3} display the results for $\alpha_S$ used in our analysis for different values of $Q^2$. We also show, on the right, the extracted
value of the effective $\alpha_S$ from the GDH sum rule. We therefore suggest large $x$ evolution in DIS as yet another way of defining  an effective coupling constant at low values of the 
scale. A more quantitative analysis to relate different types of measurements, and to study in depth the possible process dependence of $\alpha_S$ is in 
progress \cite{ChenDeurLiu}.   

\vspace{0.3cm}
In conclusion, we believe there is a much richer structure to the scale dependence of the nucleon's distribution functions
that persists
behind the apparent cancellation among higher twist terms.
We started uncovering this structure in the initial work of Refs.\cite{Kep1,BFL}. 
While on one side this points at the fact that PQCD provides an essential framework for understanding the working of duality, 
on the other a thorough understanding of the lack of final state interactions  is still missing.  
Our analysis opens up the possibility of extracting 
the effective strong coupling constant, $\alpha_S$, at low scale
from a different process than in Refs.\cite{ChenDeur,Deur1}.

\vspace{0.3cm}
This work is supported by  the U.S. Department
of Energy grant no. DE-FG02-01ER41200. 



\bibliographystyle{aipproc}   




\end{document}